# Likelihood Based Study Designs for

# Time-to-Event Endpoints

**Jeffrey D. Blume\***

Department of Biostatistics, Vanderbilt University School of Medicine, Nashville, Tennessee, 37232 USA

*\*email:* j.blume@vanderbilt.edu

**and**

**Leena Choi\***

Department of Biostatistics, Vanderbilt University School of Medicine, Nashville, Tennessee, 37232 USA

*\*email:* leena.choi@vanderbilt.edu

SUMMARY:    Likelihood methods for measuring statistical evidence obey the likelihood principle and maintain excellent frequency properties. These methods lend themselves to sequential study designs because they measure the strength of statistical evidence in accumulating data without needing adjustments for the number of planned or unplanned examinations of data. However, sample size projections have, to date, only been developed for fixed sample size designs. In this paper, we consider sequential study designs for time-to-event outcomes assuming likelihood methods will be used to monitor the strength of statistical evidence for efficacy and futility. We develop sample size projections with the aim of controlling the probability of observing misleading evidence under the null and alternative hypotheses, and we show how efficacy and futility considerations are managed in this context. We also consider relaxing the requirement of specifying the simple alternative hypothesis in advance of the study. Finally, we end with a comparative illustration of these methods in a phase II cancer clinical trial that previously was designed within a Bayesian framework.

## 1. Introduction

When it comes to measuring the strength of statistical evidence in a given body of data, likelihood ratios are ideally suited to the task. They do not depend on the sample space or the prior distribution, and they retain excellent frequency properties despite strictly adhering to the likelihood principle. Because of this, likelihood ratios are particularly useful tools for measuring the strength of statistical evidence in sequential trials, where planned and unplanned examinations of the data can make tail area computations cumbersome and controversial.

In an effort to make likelihood methods more practicable, we provide likelihood based formulae for sample size projections in sequential trials with time-to-event endpoints. Not surprisingly, the underlying mathematical landscape is similar to that of sequential hypothesis testing. However, there are subtle and important mathematical differences, e.g., likelihood methods hold the criteria for strong evidence constant as the sample size grows instead of holding the Type I error rate constant. Our aim is to present a comprehensive framework for evaluating the frequency properties of likelihood sequential study designs. This includes identifying key concepts, establishing terminology, and deriving formulae for the probabilities of observing misleading and strong evidence, and the expected sample size in a given study design.

The operational characteristics of likelihood study designs depend on two simple pre-specified null and alternative hypotheses of interest. The resulting likelihood ratio, the measure of the strength of evidence, is often criticized for its dependence on a pre-specified simple alternative hypothesis. Although we do not agree with this criticism, we explore what happens when the pre-specified alternative is replaced by the best supported post-hoc alternative. As expected, the chance of being led astray does increase. However, it does not increase as much as one might expect. Moreover, we offer a very simple design modification that keeps this probability to acceptable levels: forgo very early examination of the data, e.g., forgo sequential examination until 10% of the data are collected. The ability of likelihood ratios to reliably accommodate post-hoc alternatives, if needed, should eliminate concerns about pre-specifying a simple hypothesis.

We first provide a brief review of key likelihood concepts. Those familiar with likelihood methods for measuring statistical evidence can skip this section. We then consider relaxing the dependence of likelihood methods on a fixed simple alternative by considering post-hoc alternatives. We then go on to consider sequential designs of time-to-event outcomes, develop formulae for sample size projections, and discuss interim projections and futility assessments. Finally, we illustrate these methods with a real example from a (confidential) Phase II cancer trial of survival that was designed using Bayesian methods. Throughout this paper the parameter of interest will be the hazard ratio comparing the experimental hazard to the control hazard. As such, hazard ratios less than one indicate that the experimental therapy has a smaller hazard and tends to produce longer survival times.

## 2. Likelihood Preliminaries

The law of likelihood explains when data represent statistical evidence for one hypothesis over another. In particular, data better support the hypothesis that does a better job of predicting the observed events and the likelihood ratio measures the degree to which one hypothesis is better supported over another (Hacking, 1965; Edwards, 1972; Royall, 1997). Introductory material on the Law of Likelihood, recent applications, and technical advances (e.g., robustness) are available in the literature (Blume, 2002; Goodman and Royall, 2002; Blume, 2005; Blume et. al., 2007; Blume, 2008; Strug and Hodge, 2006a,b; Royall and Tsou, 2003; Van der Tweel, 2005; Choi et. al., 2008). Sample size projections for fixed sample size designs is discussed in Royall (1997) and Strug et. al. (2007). Various likelihood methods for clinical trials can be found in Piantadosi's popular and often-cited text on clinical trials (Piantadosi, 2005).

The law of likelihood should not be confused with the likelihood principle, which gives the conditions under which two experiments yield equivalent statistical evidence for two hypotheses of interest (Hacking, 1965; Birnbaum, 1962; Barnard, 1949). It follows from the law that there are three evidential quantities: (1) a measure of the strength of evidence; (2) the probability that a particular study design will generate misleading evidence; and (3) the probability that observed evidence is misleading (Blume, 2008). Many of Likelihood's favorable attributes arise because of this distinction .

### 2.1 *Notation and Setup*

Define the log hazard ratio to be $\theta = \ln(\lambda_t/\lambda_c) = \ln\psi$ where ($\lambda_t$ is the hazard rate for the treatment, $\lambda_c$ is the hazard rate for the control group, and $\psi$ is their hazard ratio. We will use Cox's Partial Likelihood (Cox, 1972, 1975) to characterize the evidence about the log hazard ratio

$$L(\theta) = \prod_{i=1}^{n} \left( \frac{\exp\{\theta Z_i\}}{\sum_{j \in R_i} \exp\{\theta Z_j\}} \right)^{\delta_i} \tag{1}$$

where $Z_i$ is an indicator variable for treatment, $\delta_i$ is a censoring indicator (zero if the $i^{th}$ subject is right censored and one otherwise), and $R_i$ is the risk set of subjects immediately prior to the $i^{th}$ event. Likelihood functions are invariant to parameter transformations, so the evidence about $\theta$ and $\exp\{\theta\}$ is identical.

For two competing hypotheses of interest, say $H_0 : \theta = \theta_0$ and $H_1 : \theta = \theta_1$, the strength of the evidence for $H_1$ over $H_0$ is measured by $LR = L(\theta_1)/L(\theta_0)$[1]. As the sample size grows, the (partial) likelihood ratio will converge to either 0 or $\infty$ in support of the true hypothesis.

---

[1]Likelihood ratios less than 8 indicate 'weak' evidence and likelihood ratios greater than 32 indicate 'strong' evidence. A likelihood ratio of 20 is well characterized as 'fairly strong' or 'moderate' evidence.

The observed likelihood ratio will fall into one of three regions: $LR \in [0, 1/k]$ indicating strong evidence for $H_0$ over $H_1$, $LR \in (1/k, k)$ indicating weak evidence, and $LR \in [k, \infty)$ indicating strong evidence for $H_1$ over $H_0$. Conventional benchmarks of $k = 8, 32$ are points of reference along the gradual shift from weak to moderate evidence ($k = 8$) and from moderate to strong evidence ($k = 32$). The collection of parameter values that are best supported by the data at the $k^{th}$ evidential level is called a $1/k$ likelihood support interval (Blume, 2002; Royall, 1997).

### 2.2 *Misleading Evidence*

Misleading evidence is defined as strong evidence in favor of the incorrect hypothesis (e.g., observing $LR > k$ when $H_0$ is true). We never know if observed evidence is misleading or not; all we can report is the strength of the observed evidence. As a result, it is important to understand how often likelihood ratios can be misleading. There are a number of important results that show that the probability of observing misleading evidence is low and controllable. We very briefly mention these results here because they will be useful in later sections.

#### 2.2.1 *The Universal Bound.*   For any fixed sample size, the probability of observing misleading evidence is bounded above by the inverse of the strength of evidence.

$$mis_0 = P_0 \left( \frac{L(\theta_1)}{L(\theta_0)} \geqslant k \right) \leqslant \frac{1}{k} \qquad (2)$$

where $1/k$ is the so-called Universal bound. This bound applies when comparing any two hypotheses where the probability of observing the data is well defined (Pratt, 1977; Royall, 1997) and, more generally, to partial likelihoods such as (1) (Eddings, 2003; Huang, 1998).

#### 2.2.2 *The Bump Function.*   As the sample size grows, the probability of observing misleading evidence converges to zero (i.e., $mis_0 \to 0$ as $n \to \infty$). This is a consequence of the fact that a likelihood ratio will converge to either 0 or $\infty$ in support of the true hypothesis.

In large samples, the probability of observing misleading evidence is virtually nonexistent and not a concern for the experimenter. In our semi-parametric setting, a good large sample approximation to the probability of observing misleading evidence is

$$mis_0 \approx \Phi[-\ln k/\Delta\sqrt{n} - \Delta\sqrt{n}/2] \tag{3}$$

where $\Delta$ is the distance between the hypotheses in expected information units, $\Delta = |\theta_1 - \theta_0|\sqrt{I(\theta_0)}$ (Eddings, 2003; Royall, 2000). This approximation is based on the asymptotic normality of the score function and is exact, for any sample size, when the underlying model is $N(\mu, \sigma^2)$ so that $\Delta = |\mu_1 - \mu_0|/\sigma$. Figure 1 displays (3) under that normal model when $n = 3$ (curve 'a') and $n = 15$ (curve 'b').

[Figure 1 about here.]

This approximation to $mis_0$ is called the Bump function (Royall, 2000) because of its symmetrical appearance. This function represents the probability of observing misleading evidence for an alternative $\Delta$ information units away from the null hypothesis on the $n^{th}$ observation. A key observation is that the maximum probability of observing misleading evidence, over all alternatives, is $\Phi[-\sqrt{2\ln k}]$. When $k \geqslant 8$ this maximum tends to be much less than the universal bound. The maximum is achieved at alternatives that are $\pm\sqrt{2\ln k}$ standard errors from the null hypothesis.

2.2.3 *The Tepee and Extended Bump Function.* In a sequential trial the data are examined after each observation is collected so the Bump function no longer describes the probability of observing misleading evidence. Specifically, the data are examined beginning with the $m_0^{th}$ observation and continuing until the $m^{th}$ observation. The Extended Bump function gives the probability of observing misleading evidence between the $m_0^{th}$ and $m^{th}$ observation, when the data are examined after each observation in that range (Blume, 2008).

The Extended Bump function is

$$P_0\left(\frac{L(\theta_1)}{L(\theta_0)} \geqslant k \;\; ; \text{for any } n \in [m_0, m]\right) \tag{4}$$

$$\cong \Phi\left[-\left(\frac{\ln k}{\Delta} + \rho\right)m^{-\frac{1}{2}} - \frac{\Delta}{2}m^{\frac{1}{2}}\right]$$

$$+\Phi\left[-\left(\frac{\ln k}{\Delta} + \rho\right)(m_0-1)^{-\frac{1}{2}} - \frac{\Delta}{2}(m_0-1)^{\frac{1}{2}}\right]$$

$$+\frac{\exp\{-\rho\Delta\}}{k}\left\{\Phi\left[\left(\frac{\ln k}{\Delta} + \rho\right)(m_0-1)^{-\frac{1}{2}} - \frac{\Delta}{2}(m_0-1)^{\frac{1}{2}}\right]\right.$$

$$\left.-\Phi\left[\left(\frac{\ln k}{\Delta} + \rho\right)m^{-\frac{1}{2}} - \frac{\Delta}{2}m^{\frac{1}{2}}\right]\right\}$$

where $\rho$ is a model dependent constant that corrects the underlying continuous time approximation for use in discrete time. Details can be found in (Blume, 2008) and the references therein. This approximation works in information time $(I(\theta))$ as opposed to participant time $(n)$. This approximation is very accurate when the underlying model is $N(\mu, \sigma^2)$, in which case we have $\Delta = |\mu_1 - \mu_0|/\sigma$ and $\rho = 0.583$.

Setting $m_0 = 1$ and letting $m \to \infty$ yields an important special case. The result is the Tepee function (Blume, 2008), which represents the maximum probability of observing misleading evidence in a sequential study. The Tepee function is

$$P_0\left(\frac{L(\theta_1)}{L(\theta_0)} \geqslant k \;\; ; \text{for any } n = 1, 2, \ldots\right) \approx \frac{\exp\{-\rho\Delta\}}{k} \leqslant \frac{1}{k} \tag{5}$$

Assuming a normal model, figure 1 displays the Tepee function (curve 'c') and the Extended Bump function (curve 'd') when $m_0 = 3$ and $m = 15$. Note that despite an infinite number of looks at the data (one after each observation) the probability remains bounded by the universal bound $1/k$.

The Extended Bump function provides a useful approximation to the probability of observing misleading evidence because investigators are rarely willing to stop a study very early and funding always limits the total number of observations. The Tepee function gives the worst case-scenario for observing misleading evidence. These scenarios mimic the behavior of an unscrupulous investigator who ignores all evidence not supporting his pet hypothesis and continue to collect observations until his pet alternative is better supported over the null hypothesis. Fortunately, the unscrupulous investigator will be thwarted at least $(1-1/k)100\%$ of the time. Study designs that are not as biased towards the pre-specified alternative have a smaller potential to generate misleading evidence and study designs that replace the pre-specified alternative with a post-hoc one will have a larger potential to generate misleading evidence.

## 3. Accommodating post-hoc alternatives

There is always the temptation to examine the strength of evidence for alternatives not originally pre-specified. And because the alternative hypothesis is explicit in likelihood methods, it is important to consider how often this activity would lead investigators astray. As we show in the next two sub-sections, the answer is "not very often". We find that a price is paid for using a post-hoc alternative; the data are more likely to favor some false alternative - as opposed to the pre-specified alternative - over a true null hypothesis. However, this increase is negligible in fixed sample size designs, and small and manageable in practical sequential designs.

A note on terminology: we will refer to the probability of finding some false (post-hoc) alternative that is better supported over the true null hypothesis as *the probability of being led astray*. We do this to distinguish it from the probability of observing misleading evidence, which depends on two fixed (pre-specified) simple hypotheses and is represented by the Bump, Tepee and Extended Bump functions.

Recall in our survival analysis example that $\theta$ is the log hazard ratio for treatment over control. Any $\theta < 0$ indicates that the treatment is beneficial (smaller hazard) to some degree. Mathematically, there is strong evidence in the data supporting some alternative (some $\theta < 0$) over the null hypothesis ($H_0 : \theta = 0$) whenever $\{\sup_{\theta<0} L(\theta)/L(0) \geqslant k\}$ where

$$\sup_{\theta < 0} \frac{L(\theta)}{L(0)} = \begin{cases} L(\widehat{\theta})/L(0) & \text{if } \widehat{\theta} < 0 \\ \\ 1 & \text{if } \widehat{\theta} \geqslant 0 \end{cases} \tag{6}$$

where $\widehat{\theta}$ is the MLE. For the general null hypothesis $H_0 : \theta = \theta_0$ we write this condition as $\{\sup_{\theta < \theta_0} L(\theta)/L(\theta_0) \geqslant k\}$. The probability of being led astray is $P_0 \left( \sup_{\theta < \theta_0} L(\theta)/L(\theta_0) \geqslant k \right)$.

Note that we are careful to avoid interpreting $\sup_{\theta < 0} L(\theta)/L(\theta_0)$ as a evidential likelihood ratio for the composite hypothesis $H_c : \theta < 0$ over $H_0$. This is because, according to the Law of Likelihood, it is not. The generic problem is that $\sup_{\theta < 0} L(\theta)$ does not constitute a properly specified probability distribution for the 'alternative hypothesis' (Royall, 1997; Blume, 2002). There are a number of interesting philosophical viewpoints about this, but we will not delve into them here. It suffices to simply use the expression $\sup_{\theta < 0} L(\theta)/L(\theta_0)$ to indicate when there exists at least one post-hoc alternative will be better supported over the null hypothesis.

### 3.1 *Fixed sample size designs*

In a fixed sample size design, where the examination of the data occurs only at the end of the study, the probability of being led astray is approximately $\Phi[-\sqrt{2\ln k}]$ - the maximum of the bump function. That is,

$$P_0 \left( \sup_{\theta < \theta_0} \frac{L(\theta)}{L(\theta_0)} \geqslant k \right) \approx \Phi[-\sqrt{2\ln k}] \tag{7}$$

this approximation is exact when the underlying model is normal and holds otherwise in moderately large samples under standard regularity conditions (see appendix). The driving force of this result is the asymptotic normality of the score function; direct calculation yields the result for the normal case. Because the probability of being led astray is no worse than the maximum probability of observing misleading evidence for a fixed alternative, there appears to be little penalty for evaluating post-hoc alternatives hypotheses in fixed sample size designs.

3.2 *Sequential designs*

We consider fully sequential designs where the data are examined beginning with the $m_0^{th}$ observation and continuing until the $m^{th}$ observation. These constraints on monitoring are practical: $m$ is set by the budget and $m_0$ is often a scientific minimum need for external validity. Because the accumulating data are examined repeatedly, the probability of being led astray can be much higher than in a fixed sample size design. But how much higher depends on the sample constraint ratio $m/m_0$.

The probability of being led astray in a sequential design is bounded by

$$P_0 \left( \sup_{\theta < \theta_0} \frac{L(\theta)}{L(\theta_0)} \geqslant k \ \ ; \text{for any } n \in [m_0, m] \right) \leqslant \frac{\sqrt{\ln k}}{2k\sqrt{\pi}} \ln \left( \frac{m}{m_0} \right) \qquad (8)$$

See appendix for a sketch of the proof. This bound depends on the (often very good) Brownian motion approximation to the score function and is exact for normal models under standard conditions. This bound is not necessarily achievable and its degree of 'tightness' depends on the constraint ratio $m/m_0$ and $k$. Simulation studies can be used to provide a more precise approximation[2].

When $m_0 = 1$ and $m = \infty$, the probability of being led astray, for any fixed $k$, is one (this is a consequence of the law of the iterated logarithm). This should not be surprising; it will always be possible to find strong evidence for some (unspecified) alternative - imaging one infinitely close to the null hypothesis - at some point in time when we are allowed to look forever[3]. But what may be surprising is that this probability is much lower for very practical sample constraint ratios $m/m_0$.

[Table 1 about here.]

Table 1 displays the probability observing at least $k$-strength evidence for some false alternative over the true null (i.e., the probability of being led astray) when examination of the data begins after collecting the $m_0^{th}$ observation and continues repeatedly until the $m^{th}$ observation has been collected. This is a very realistic scenario for medical research; budget constrains set $m$ and constraints on experimental validity often set $m_0$. The table shows, perhaps contrary to intuition, that the probability of being led astray is sufficiently low and controllable through the sample constraint ratio $m_0/m$.

---

[2]We have considered only one-sided alternative hypotheses, but the results extend to two sided alternatives by doubling the bound.

[3]If this were not the case, then we would only need to collect a few observations to discover the true hypothesis.

An illustration is helpful. Suppose the analysis plan for a study called for repeatedly examining the data after enrolling 10% of the maximum allowable sample size. The plan also calls for stopping the study as soon as strong evidence for *any* hypothesis indicating $\theta < 0$ is better supported over the null hypothesis by a factor of 20 or more. Now, if the null hypothesis is true, then by Table 1 we see that the probability of being led astray is only 0.0562. For a trial with 500 participants this amounts to 450 looks at the data, where the alternative is left unspecified and essentially reset by the data after each observation. Yet we will only be lead astray 5.6% of the time when the null hypothesis is true. The take home message is that the evaluation of post-hoc alternatives hypotheses is not terribly misleading in this context.

## 4. Sequential Survival Designs

In this section we detail the operational characteristics of a sequential design that uses likelihood methods to measure the strength of evidence in the data. Recall our parameter of interest is the log hazard ratio $\theta$. The planned goal of our study is to generate strong evidence for one of the two pre-specified simple hypotheses of interest, say $H_1$ and $H_0$, over other under. When dealing with time-to-event endpoints the Cox partial likelihood is an attractive option in the analysis stage, but less so for planning purposes because it lacks the full analytical form need to yield complete projections. Therefore we consider instead two approaches for projecting the sample size: the first is based on the normal approximation to the log hazard ratio and the second is based on an underlying Poisson approximation.

### 4.1 *Projections Based on Normality Approximations*

A well known normal approximation to the distribution of the log hazard ratio $\theta$ is

$$\widehat{\theta} \sim N\left(\theta, 4/d\right) \tag{9}$$

where $d$ is the total number of events in both groups and $\psi = \lambda_t/\lambda_c = \exp\{\theta\}$ is the hazard ratio. Note that if $d_i$ represents the number of events in the $i^{th}$ group then the variance is $1/d_t + 1/d_c$ which is $\approx 4/d$ when $d_t \approx d_c$ and $d = 2d_t$. This form of the variance is useful for planning purposes when the number of events in the two groups is expected to be disparate.

Let $\Delta = |\theta_1 - \theta_0|/2$. If $\psi_1 = 0.415$, then $\theta_1 = -0.8795$ and $\Delta = 0.44$. If $\psi_1 = 0.61$, then $\theta_1 = -0.4943$ and $\Delta = 0.25$. The sample size projections are driven by number of events and not the number of participants. This should be clear from the variance in (9), although $\widehat{\theta}$ may change along with exposure time.

A slightly more accurate approximation to the expected value and variance of the log-rank test statistic can be used to improve the projections. Using the fact that the log-rank statistic is approximately $\sqrt{d/4}\ln\psi$, and Freedman's (Freedman, 1982) non-parametric formulation for the mean and variance of the log rank statistic, we have that

$$\widehat{\theta} \sim N\left(\frac{2(\psi-1)}{(\psi+1)}, \frac{16\psi}{d(\psi+1)^2}\right) \tag{10}$$

To see the connection with (9) notice that $2(\psi-1)/(\psi+1) \approx \ln\psi = \theta$ due to a first order power series expansion and the variance reduces to $4/D$ when $2\sqrt{\psi}/(\psi+1) \approx 1$. As a result, we can see that (9) works well under the null hypothesis that $\psi = 1$ and less so as $\psi$ moves away from 1. Freedman finds that (9) is conservative when the alternative is true but not fatally so (Freedman, 1982).

## 4.2 *Operational Characteristics*

The trial will be stopped if there is $k_1$-strength evidence in favor of the alternative hypothesis $H_1 : \theta = \theta_1$ or $k_0$-strength evidence in favor of the null hypothesis $H_0 : \theta = \theta_0$ for some $k_0 < 1 < k_1$. That is, the trial continues as long as there is weak evidence such that $k_0 \leqslant L_d(\theta_1)/L_d(\theta_0) \leqslant k_1$ where $d$ is the total number of observed events. If $L_d(\theta_1)/L_d(\theta_0) > k_1$, then the trial stops for efficacy; if $L_d(\theta_1)/L_d(\theta_0) < k_0$ then the trial stops for inefficacy (i.e., the null is better supported over the alternative)[4].

There are two likelihood rates of misleading evidence that are analogous to the Type I and Type II error rates of hypothesis testing: let $\alpha_l = P_0 \left( L_d(\theta_1)/L_d(\theta_0) > k_1 \right)$ and $\beta_l = P_1 \left( L_d(\theta_0)/L_d(\theta_1) > 1/k_0 \right)$ [5]. For simplicity, we ignore any practical limits on the sample size; we will return to this issue later. Let $b = \ln(k_1)/\Delta$ and $a = \ln(k_0)/\Delta$, so that when $k_1 = 1/k_0 = k$ we have $b = -a$. Remember that $\Delta = |\theta_1 - \theta_0|/\sigma$ because of our normality assumption for the log hazard ratio. Then it follows from Blume's (Blume, 2008) likelihood translation of Siegmund's (Siegmund, 1985) and Wald's (Wald, 1947) seminal works in sequential analysis that:

$$
\begin{aligned}
\alpha_l &= \frac{1 - e^{(a-\rho)\Delta}}{e^{(b+\rho)\Delta} - e^{(a-\rho)\Delta}} \leqslant \frac{1 - k_0}{k_1 - k_0} \\
\text{power}_l = 1 - \beta_l &= \frac{1 - e^{-(a-\rho)\Delta}}{e^{-(b+\rho)\Delta} - e^{-(a-\rho)\Delta}} \geqslant k_1 \frac{1 - k_0}{k_1 - k_0}
\end{aligned}
\tag{11}
$$

With the following expected number of events:

$$
\begin{aligned}
E_0[D] &= \frac{(b+\rho)\,\alpha_l + (a - \rho)\,(1 - \alpha_l)}{\phi_0} \\
E_1[D] &= \frac{(b+\rho)\,(1 - \beta_l) + (a - \rho)\,\beta_l}{\phi_1}
\end{aligned}
\tag{12}
$$

---

[4]Inefficacy and futility are not the same concept. Inefficacy is discovering evidence for the null hypothesis over the alternative (e.g., the treatment does not work), while futility is the failure to generate sufficient evidence for either efficacy or inefficacy.

[5]This definition may appear to include weak evidence in the definition of power. However this is not the case because, for example, $1 - \beta_l = P_1(L_1/L_0 > 1/8 \text{ and } L_1/L_0 > 8) + P_1(L_1/L_0 > 1/8 \text{ and } L_1/L_0 < 8)$ and the last expression is identically zero in a limitless sequential trial.

where $\phi_0 = -\Delta/2$, $\phi_1 = \Delta/2$ and $\rho$ is a constant that accounts for the expected overshoot of the stopping boundary in discrete time[6]. Wald proved that under either hypothesis, the probability of eventually finding strong evidence in either direction is one as long as it is possible for the sample size to be very large (Wald, 1947). Simulations (not detailed here) indicate these approximations are accurate as long as the practical upper limit on the sample size is in the upper quantiles of the stopping time, say 90% or 95%.

The trade-off among the evidential levels and the operating characteristics are displayed in Tables 2 and 3 for a variety of different study designs with $\Delta = 0.44, 0.25$ ($\psi_1 = 0.415, 0.61$, these values were chosen based on our upcoming example). The centiles of the stopping time distribution (i.e., number of events at which the study would stop) were obtained by simulation. As the desired level of evidence increases, so too does the expected number of events. Notice that as the the strength of evidence increases, $1 - \beta_l$ increases and $\alpha_l$ decreases. This happens because stray or misleading sample paths have plenty of time to correct themselves.

[Table 2 about here.]

[Table 3 about here.]

Remember that this is an approximation in information time; the number of events must be translated into the number of participants using the relation $n = d/p$, where $p$ is the expected proportion of participants who will have the event in the study time frame. For example, a study needing between 25 and 55 events would have to enroll between 32 (=25/0.8) and 69 (=55/0.8) participants if 80% of participants are expected to have the event during the study follow-up period.

---

[6]Typically $\rho = 0.583$ for the normal distribution and $\rho = 0.32$ for the binomial distribution (Siegmund, 1985).

### 4.3 *Projections Based on Poisson Assumptions*

Let $d_i$ be the number of events and $t_i$ be the exposure time in the $i^{th}$ group ($i = t, c$; treatment or control). Here we assume $d_c \sim Poiss(\lambda_c t_c)$ and $d_t \sim Poiss(\lambda_t t_t)$ as in (Berry, 1983; Royall, 1997). With $d_c + d_t = d$ we have the conditional distribution $d_t | d \sim \text{Bin}(d, p)$ where $p = \lambda_t t_t / (\lambda_c t_c + \lambda_t t_t) = \psi / (\psi + g)$ with hazard ratio $\psi = \lambda_t / \lambda_c = \exp\{\theta\}$ and exposure ratio $g = t_c / t_t$. The conditional likelihood function for the hazard ratio, $\psi$, is then

$$L(\psi | d, g) \propto \left( \frac{\psi}{\psi + g} \right)^{d_t} \left( \frac{g}{g + \psi} \right)^{d_c} \tag{13}$$

We can use the likelihood function's invariance to reparameterization and base our sample size projections on the binomial distribution[7]. This works because for fixed $g$, we have a simple mapping from $H_i : \psi = \psi_i$ to $H_i : p = p_i$ by way of $p_i = \psi_i / (\psi_i + g)$ for $i = 0, 1$.

If we let $\Delta = \ln[p_1(1 - p_0)/(1 - p_1)p_0] = \ln[\psi_1 / \psi_0]$ and $\rho = 0.32$ and $\phi_i = p_i + \ln[(1 - p_1)/(1 - p_0)]/\Delta$, then equations (11) and (12) apply whenever $p_1 > p_0$. If $p_1 < p_0$ - as it is in our example because $0.415 = \psi_1 < \psi_0 = 1$ - we can reparameterize to the symmetrical equivalent of $\psi_1 = 1/0.415 = 2.41$, where $\psi$ is now the hazard ratio for control to treatment.

[Table 4 about here.]

Table 6 displays the expected number of events, $E_i[D]$, and the operating characteristics, $\alpha_l$ and $1 - \beta_l$, using this approach. It is reassuring to see that Table 6 is in good agreement with the projections based on the normal approximation to the log hazard given in Table 2. Note that quantiles obtained by simulation in Table 6 assume $\lambda_c = 0.25$. Agreement on the frequency characteristics and expected values is to be expected when $g = 1$, but agreement on stopping time quantiles is sensitive to specification of $\lambda_c = 0.25$[8]. When $g \neq 1$, the alternative formulation for the variance, $1/d_t + 1/d_c$, should be used in (9) if the two approaches are to agree.

---

[7] This likelihood can also be used for the analysis.

[8] Although our simulations and experience suggest it is not overly sensitive.

Under this model, we can project the exposure time needed to obtain the desired number of events, say $d^9$, with probability at least $\gamma$, say 0.8. Assuming the ratio of exposure times and hazards remains constant, we have $D \sim Poiss(\lambda_i)$ where $\lambda_i = \lambda_c t_c(1 + \psi_i/g)$ for $i = 0, 1$ which depends on the null and alternative hypotheses $\psi_i$. Then, with knowledge of $\lambda_c$, it is a matter of numerical calculation to find the smallest $t_c$ such that

$$P_i(D \geqslant d) = \sum_{j=d}^{\infty} \frac{e^{-\lambda_i}(\lambda_i)^j}{j!} \geqslant \gamma \tag{14}$$

for $i = 0, 1$ and from $t_c$ we obtain $t_t = t_c/g$. An alternative to this numerical search would be the following projection

$$t_c = \frac{E_i[D]}{\lambda_c(1 + \psi_i/g)} \tag{15}$$

for $i = 0, 1$, which comes from $\lambda_i = E_i[D]$.

## 5. Illustration

This work was motivated by the authors' desire to provide an alternative design to an ongoing Phase II cancer clinical trial of a chemotherapy agent (say drug X) in participants undergoing resection for pancreatic cancer[10]. Recurrence-free survival was the primary endpoint and all-cause mortality was the secondary endpoint, so time-to-event methods were warranted. The study was originally designed from a Bayesian perspective using data previously published in (Oettle et. al., 2007). The simulations shown in this section use those published data on survival rates as if they represented the true population. This provides a welcome opportunity to assess the impact of the normality assumptions upon which our likelihood projections are based.

---

[9] $d$ need not be the expected value; e.g., it could be the 90% of the stopping time distribution.

[10] The trial is ongoing and we must maintain confidentiality regarding the active drug and the sponsor.

## 5.1 *Bayesian design*

Previous studies yielded Kaplan-Meier estimates of disease-free survival and these data were used to anticipate the hazard rate for the new treatment versus standard therapy: $H_1$ : $\psi = 0.415$ ($\Delta = 0.44$). Several different survival models were considered, but the cox model appeared to fit as well as any other model.

In the trial being designed, participants would be recruited over a 2.4 year period and randomly assigned to the new treatment agent or standard therapy. A maximum of fifty patients per group were to be recruited and participants could be followed for a maximum for 4.5 years. As designed, the trial would be stopped as soon as the posterior probability of the log hazard ratio being less than 0 was greater than 0.95 (i.e., $P(\theta < 0) > 0.95$). A normal skeptical prior for the log hazard ratio with mean of 0 and standard deviation of 0.5606 was used because the probability of the hazard ratio being greater than 3 was 0.025[11]. The posterior distribution of the log hazard ratio would be examined as soon as 10 events were observed and every event thereafter.

We also considered a slightly modified design where the trial would be stopped if the posterior probability of the log hazard ratio being less than 0 was either greater than 0.95 or less than 0.1 (i.e., not $0.1 < p(\theta < 0) < 0.95$). Here too, we had a maximum of fifty patients per group and the data would not be examined until 10 events were observed.

Monte-Carlo simulation was used to determine the operational characteristics of the initial and modified designs. Interestingly, both designs yielded the same operational characteristics with the exception of the stopping time distribution. The probability of incorrectly stopping the trial when the null is true ($\psi = 1$) was 0.132 (Type I error rate); the probability of not stopping under the null was 0.654. Also, the probability of correctly stopping the trial when

---

the alternative was true ($\psi = 0.415$) was 0.96; the probability of not stopping under the alternative was 0.038.

Increasing the cap on the sample size from 50 to 1000 did not yield any differences beyond those in the stopping time distribution between the initial design and its modified version. With a cap of 1000, the probability of incorrectly stopping the trial when the null is true ($\psi = 1$) was 0.322 (Type I error rate); The probability of not stopping under the null was 0.134. Also, the probability of correctly stopping the trial when the alternative was true was 0.996; the probability of not stopping under the alternative was essentially 0. Clearly, the prior is playing a noticeable role and further investigations are warranted.

## 5.2 *Likelihood design*

Table 6 shows the simulated operating characteristics of the likelihood based design. Notice that there is an additional column on this table labeled "non-stop" that counts the proportion of simulations where strong evidence was not observed in either direction before resources ran out, resulting in an inclusive study (i.e. a study that generated only weak evidence). If the maximum number of events is increased to only 130, the resulting simulations (not shown here) are virtually identical to the theoretical projections in table 2.

[Table 5 about here.]

To permit comparison between the operation characteristics of the Bayesian and likelihood designs, their stopping criteria must be calibrated. This is done by noting that the Bayes factor, equal to the posterior odds divided by the prior odds, is also a simple vs. simple likelihood ratio under the full model with priors included. So the stopping criteria $0.1 < p(\theta < 0) < 0.95$ translates to $1/9 < LR < 19$ because the prior odds are one. Here the $K_0 = 20$ and $k_1 = 20$ is the closest to the criterion we are looking for. The probability that the likelihood approach ends without generating strong evidence is 0.056 under the null and 0.112 under the alternative. The comparable Type I error rate is 0.032 and power is 0.856.

Overall, the likelihood approach fairs well when compared to the Bayesian design with this particular prior. Note also that the likelihood design is at a slight disadvantage when compared to the Bayesian design, because the prior that was used essentially assumes knowledge of and additional 13 previous events ($d = 4/(0.5606)^2$). This increase in precision is largely responsible for the increased power with the Bayesian design and the increase in the probability of not stopping under the null hypothesis.

### 5.3 Synopsis of the likelihood design

Our study has two initial hypothesis of interest: a null hypothesis that states the hazard ratio is 1 and an alternative hypothesis that states the hazard ratio is 0.415. It is desirable to stop the study as soon as strong evidence in favor of either hypothesis over the other is found. The strength of the evidence in the data will be measured with a likelihood ratio and likelihood ratios greater than 20 will indicate strong enough evidence to stop the trial.

Consider a fully sequential design where the data are examined after each observation is collected and the study is stopped as soon as strong evidence for one of the hypotheses (i.e., null or alternative) over the other is found. In a sequential design the sample size, or stopping time, is random. With the aforementioned hypotheses, this design has an average stopping time of 32 total events[12] with a median stopping time of 25 events and $80^{th}$- and $95^{th}$-percentiles of 45 and 75 events[13]. Thus it is reasonable to plan on needing no more than 75 events. If 20% of the participants are expected to remain event free, then no more than 94 participants would be required. However, keep in mind that 50% of the time strong evidence will be observed with less than 32 participants ($32 = 25/0.8$).

The operating characteristics of this fully sequential design are as follows. If the null hypothesis is true, the probability of observing strong evidence for the alternative - misleading evidence - is only 0.037. If the alternative hypothesis is true, the probability of observing strong evidence for the alternative is 0.963[14]. Thus the accumulating data can be regularly examined for strong evidence supporting either pre-specified hypothesis over the other; the chance of observing misleading evidence remains low.

---

[12]Total events is the sum of the events from both groups.

[13]Here the stopping time distribution is the same under the null or alternative hypothesis because the stopping strength of evidence is the same in either direction, i.e., $LR = 20$.

There likely will be considerable interest in evaluating the evidence for other alternative hypotheses that indicate the hazard ratio is less than one. If this is done at the end of the study, then there is only a 0.72% chance of observing strong evidence for *any* alternative specifying that the hazard ratio is less than one over the null hypothesis when the null hypothesis is true (table 1). That is, there is less than 1% chance of being led astray, even when the evidence for post-hoc alternatives hypotheses is considered at the end of the study.

Moreover, post-hoc alternatives can be considered while the trial is ongoing. Suppose the trial is allowed to stop when strong evidence for *any* alternative hypothesis that indicated the hazard ratio is less than one is better supported over the null hypothesis. If the trial cannot collect more than 100 events and the null hypothesis is true, then we would be misled less than 11.24% of the time. That is, the probability of observing strong evidence for any false alternative indicating that the experimental therapy is better, when in truth the therapies are equally good, is less than 0.1124[15]. However, if we used a delayed sequential design that does not allow the data to be examined until 10 events have been observed[16], then the probability of being misled drops to a reasonable 5.62% or less[17]. It is re-assuring that likelihood ratios remain reliable in this setting.

---

[14]This calculation assumes a potentially infinite sample size; otherwise it is possible to end the study with only weak evidence. However the probability of stopping with weak evidence is very small when the realistic limit on the sample size is greater than the $95^{th}$-percentile, as is likely here.

[15]Simulations suggest that the actual probability is approximately 0.071.

[16]Determining the number of 'run-in' participants is not easy. One must balance the scientific and statistical gains against the ethical imperatives to do no harm and to not engage in a demonstration trial.

[17]Simulations suggest that the actual probability is approximately 0.051.

## 6. Remarks

Before finishing we would like to point out the flexibility of this approach for interim projections and futility assessments. During the course of a trial it can be helpful to compute the conditional probability that the trial will eventually yield strong evidence for the alternative or null hypothesis given the observed evidence collected so far. Futility is a consideration here, as the degree to which the interim results will change, if the trial is allowed to continue, will influence the decision to continue the trial.

With likelihood methods the calculation of these conditional probabilities, or interim projections, is straightforward. Let $LR_{int}$ be the likelihood ratio at the interim analysis, $LR_{aft}$ be the likelihood ratio from the data collected after the interim analysis, and let $LR_{fin}$ be the final end-of-study likelihood ratio. Clearly, $LR_{int} * LR_{aft} = LR_{fin}$. Then

$$P_i[LR_{fin} > k | LR_{int} = k_{int}] = P_i[LR_{aft} > k/k_{int}] \qquad (16)$$

where $i = 0, 1$ and $k_{int}$ is the observed strength of evidence at the interim analysis. So, for example, after observing a likelihood ratio of 10 in an interim analysis with only 50 participants out of the planned 100, the probability of observing a final likelihood ratio greater than 32 is just $P_i(LR > 3.2$ in 50 participants) for $i = 1, 0$. Equations (3), (5), and (11) are then used to perform the desired calculations assuming a maximum of 50 participants.

In summary, we have shown that despite their lack of adjustment for multiple looks at the data likelihood ratios remain reliable measures of the strength of statistical evidence. In addition, while specification of a simple alternative hypothesis is important for determining the operational characteristics of the study design, post-hoc alternatives can be reliably examined at the conclusion of the study and even during the study if the study is designed appropriately (i.e., the data are not examined very early on). Likelihood methods for measur-ing statistical evidence have a number of positive attributes and good frequency properties in sequential studies is yet another.

### ACKNOWLEDGEMENTS

The authors would like to acknowledge Frank Harrell Jr. for his insightful comments on this topic and for providing access to the motivating example.

<div align="center">APPENDIX SKETCH OF PROOFS</div>

**Result (7)**:

$$P_0 \left( \sup_{\theta < \theta_0} \frac{L(\theta)}{L(\theta_0)} \geqslant k \right) \approx \Phi[-\sqrt{2 \ln k}]$$

Under a normal model, $X_1, \ldots, X_n \sim N(\theta, \sigma^2)$, direction calculation shows this relationship to be exact. Also, note that $P_0 \left( \sup_\theta L(\theta)/L(\theta_0) \geqslant k \right) = 2\Phi[-\sqrt{2 \ln k}]$. If $X_1, \ldots, X_n \sim f(X_i, \theta)$ where f is a smooth function of the real-values parameter $\theta$, then

$$\lim_{n \to \infty} P_0 \left( \sup_\theta \frac{L(\theta)}{L(\theta_0)} \geqslant k \right) = 2\Phi[-\sqrt{2 \ln k}]$$

To see this, write down the Taylor expanded difference of the log-likelihoods at the MLE around the true value $\theta_0$ and write down the score equation. Use the score equation to solve for $(\hat{\theta} - \theta_0)$ and plug this solution into the Taylor expanded difference of log-likelihoods. This yields

$$l_n(\hat{\theta}) - l_n(\theta_0) = - \left[ \frac{\partial l_n(\theta)}{\partial \theta}|_{\theta_0} \right]^2 / 2 \left[ \frac{\partial^2 l_n(\theta)}{\partial \theta^2}|_{\theta_0} \right] + R$$

Now $R = O_p(n^{-1/2})$ under mild conditions and $(1/\sqrt{n}) \partial l_n(\theta)/\partial \theta|_{\theta_0} \xrightarrow{d} N(0, I(\theta_0))$ and $(1/n)\partial^2 l_n(\theta)/\partial \theta^2|_{\theta_0} \xrightarrow{p} -I(\theta_0)$. Thus we have

$$l_n(\hat{\theta}) - l_n(\theta_0) \xrightarrow{d} |Z| \quad \text{where } Z \sim N(0,1)$$

and the result follows.

**Result (8)**:

$$P_0 \left( \sup_{\theta < \theta_0} \frac{L(\theta)}{L(\theta_0)} \geqslant k \ ; \text{ for any } n \in [m_0, m] \right) \leqslant \frac{\sqrt{\ln k}}{2k\sqrt{\pi}} \ln \left( \frac{m}{m_0} \right)$$

Let $X_1, \ldots, X_n \sim N(\theta, \sigma^2)$ and without loss of generality let $\theta_0 = 0$, $\sigma = 1$. We use a device and result from Lorden (1973). Notice that

$$P_0 \left( \sup_{\theta < \theta_1} \frac{L(\theta)}{L(\theta_0)} \geqslant k \ ; \text{ for any } n \in [m_0, m] \right) = P_0 \left( m_0 \leqslant N_1 < \frac{2 \ln k}{\Delta^2} \right) + P_0 \left( \frac{2 \ln k}{\Delta^2} \leqslant N_2 \leqslant m \right)$$

where $\Delta = |\theta_1 - \theta_0|$ and stopping time $N_1 = \inf \left\{ n : 1 \leqslant n < \frac{2 \ln k}{\Delta^2}, S_n \geqslant \sqrt{2 \ln k}\sqrt{n} \right\}$ or $\infty$ if no such $n$ exists, and stopping time $N_2 = \inf \left\{ n : \frac{2 \ln k}{\Delta^2} \leqslant n, S_n \geqslant \left[ \frac{\ln k}{\Delta} + n\frac{\Delta}{2} \right] \right\}$ or $\infty$ if no such $n$ exists. The condition in $N_1$ requires that $\Delta < \sqrt{2 \ln k/n}$, but in our case $\Delta = 0$ so this condition is always met. Therefore, $P_0 \left( \sup_{\theta < \theta_0} \frac{L(\theta)}{L(\theta_0)} \geqslant k \ ; \text{ for any } n \in [m_0, m] \right) = P_0 \left( m_0 \leqslant N_1 < \min(m, \frac{2 \ln k}{\Delta^2}) \right)$ which can be bounded by $\int_{m_0}^m (\sqrt{\ln k}/t 2k\sqrt{\pi})dt$ Lorden (1973) and this yields the result.

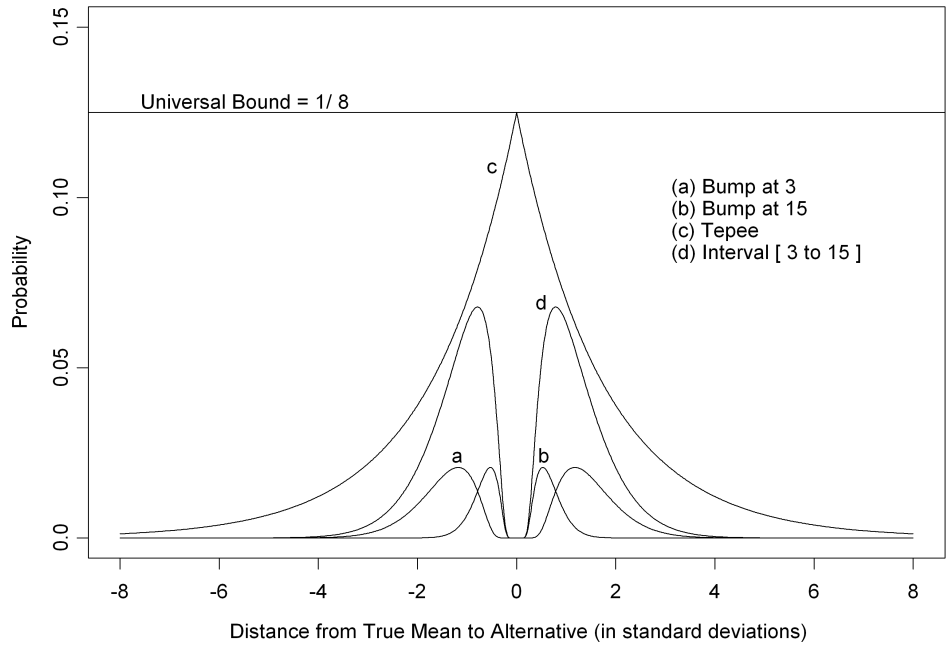

**Figure 1.** The Bump, Extended Bump, and Tepee functions

**Table 1**

*Bounds on the probability of being led astray for designs with continual examination of data after 1%, 5%, 10%, 20%, 50%, and 100% of the data have been collected. A sample constraint ratio of 1 is a fixed sample size design. Equation (8) was used to produce this table with the exception of the last column for which equation (7) was used.*

| Strength of Evidence | Sample Constraint Ratio ($m_0/m$) | | | | | |
|---|---|---|---|---|---|---|
| | 0.01 | 0.05 | 0.1 | 0.2 | 0.5 | 1 |
| 8 | 0.2342 | 0.1523 | 0.1171 | 0.0818 | 0.0352 | 0.0207 |
| 20 | 0.1124 | 0.0731 | 0.0562 | 0.0393 | 0.0169 | 0.0072 |
| 32 | 0.0756 | 0.0492 | 0.0378 | 0.0264 | 0.0114 | 0.0042 |
| 64 | 0.0414 | 0.0269 | 0.0207 | 0.0145 | 0.0062 | 0.0020 |

**Table 2**
*Likelihood study design with $H_1 : \psi = 0.415$ ($\Delta = 0.44$)*

| Evidence | | Oper. Char. | | Distribution of stopping times (events) | | | | | | |
|---|---|---|---|---|---|---|---|---|---|---|
| $k_0$ | $k_1$ | | | $E[D]$ | 25% | 50% | 75% | 80% | 90% | 95% |
| 1/8 | 8 | Null | $\alpha_l = 0.088$ | 20 | 10 | 16 | 25 | 29 | 40 | 49 |
| | | Alt | $1 - \beta_l = 0.912$ | 20 | 10 | 16 | 26 | 29 | 39 | 50 |
| 1/10 | 20 | Null | $\alpha_l = 0.036$ | 25 | 11 | 19 | 32 | 36 | 48 | 61 |
| | | Alt | $1 - \beta_l = 0.925$ | 30 | 15 | 24 | 38 | 42 | 56 | 69 |
| 1/20 | 20 | Null | $\alpha_l = 0.037$ | 32 | 16 | 25 | 41 | 45 | 61 | 75 |
| | | Alt | $1 - \beta_l = 0.963$ | 32 | 16 | 25 | 40 | 44 | 58 | 74 |
| 1/20 | 32 | Null | $\alpha_l = 0.023$ | 32 | 16 | 26 | 41 | 47 | 62 | 78 |
| | | Alt | $1 - \beta_l = 0.962$ | 36 | 19 | 30 | 46 | 51 | 67 | 82 |
| 1/32 | 32 | Null | $\alpha_l = 0.024$ | 37 | 20 | 30 | 47 | 53 | 70 | 87 |
| | | Alt | $1 - \beta_l = 0.976$ | 37 | 20 | 30 | 48 | 52 | 69 | 86 |
| 1/32 | 64 | Null | $\alpha_l = 0.012$ | 38 | 20 | 30 | 48 | 54 | 71 | 88 |
| | | Alt | $1 - \beta_l = 0.976$ | 44 | 24 | 37 | 56 | 61 | 80 | 99 |
| 1/64 | 64 | Null | $\alpha_l = 0.012$ | 45 | 25 | 38 | 57 | 63 | 82 | 100 |
| | | Alt | $1 - \beta_l = 0.988$ | 45 | 25 | 37 | 57 | 63 | 81 | 101 |

**Table 3**
*Likelihood study design with $H_1 : \psi = 0.61$ $(\Delta = 0.25)$*

| Evidence | | Oper. Char. | | Events | | | |
|---|---|---|---|---|---|---|---|
| $k_0$ | $k_1$ | | | $E[D]$ | 25% | 50% | 75% |
| 1/8 | 8 | Null | $\alpha_l = 0.098$ | 58 | 28 | 46 | 76 |
| | | Alt | $1 - \beta_l = 0.903$ | 58 | 27 | 44 | 74 |
| 1/10 | 20 | Null | $\alpha_l = 0.040$ | 72 | 33 | 54 | 93 |
| | | Alt | $1 - \beta_l = 0.917$ | 86 | 44 | 71 | 111 |
| 1/20 | 20 | Null | $\alpha_l = 0.041$ | 93 | 46 | 74 | 119 |
| | | Alt | $1 - \beta_l = 0.959$ | 93 | 46 | 74 | 119 |
| 1/20 | 32 | Null | $\alpha_l = 0.026$ | 95 | 47 | 76 | 125 |
| | | Alt | $1 - \beta_l = 0.958$ | 107 | 57 | 88 | 137 |
| 1/32 | 32 | Null | $\alpha_l = 0.026$ | 110 | 57 | 89 | 141 |
| | | Alt | $1 - \beta_l = 0.974$ | 110 | 57 | 89 | 141 |
| 1/32 | 64 | Null | $\alpha_l = 0.013$ | 113 | 57 | 90 | 142 |
| | | Alt | $1 - \beta_l = 0.973$ | 131 | 72 | 108 | 166 |
| 1/64 | 64 | Null | $\alpha_l = 0.013$ | 135 | 74 | 112 | 171 |
| | | Alt | $1 - \beta_l = 0.987$ | 135 | 74 | 111 | 168 |

**Table 4**
*Likelihood study design with $H_1 : \psi = 2.41$ and $g = 1$; Quantiles obtained by simulation assume $\lambda_c = 0.25$.*

| Evidence | | Oper. Char. | | Distribution of stopping times (events) | | | | | | |
|---|---|---|---|---|---|---|---|---|---|---|
| $k_0$ | $k_1$ | | | $E[D]$ | 25% | 50% | 75% | 80% | 90% | 95% |
| 1/8 | 8 | Null | $\alpha_l = 0.086$ | 21 | 9 | 16 | 27 | 31 | 41 | 51 |
| | | Alt | $1 - \beta_l = 0.914$ | 23 | 12 | 18 | 29 | 32 | 42 | 54 |
| 1/10 | 20 | Null | $\alpha_l = 0.035$ | 26 | 11 | 20 | 33 | 38 | 51 | 64 |
| | | Alt | $1 - \beta_l = 0.927$ | 33 | 18 | 27 | 41 | 45 | 57 | 72 |
| 1/20 | 20 | Null | $\alpha_l = 0.036$ | 33 | 16 | 26 | 42 | 47 | 62 | 77 |
| | | Alt | $1 - \beta_l = 0.964$ | 35 | 19 | 28 | 45 | 50 | 65 | 80 |
| 1/20 | 32 | Null | $\alpha_l = 0.023$ | 34 | 16 | 26 | 42 | 47 | 65 | 80 |
| | | Alt | $1 - \beta_l = 0.963$ | 40 | 22 | 33 | 49 | 54 | 72 | 89 |
| 1/32 | 32 | Null | $\alpha_l = 0.023$ | 39 | 20 | 30 | 48 | 55 | 71 | 89 |
| | | Alt | $1 - \beta_l = 0.977$ | 41 | 23 | 33 | 51 | 56 | 74 | 92 |
| 1/32 | 64 | Null | $\alpha_l = 0.012$ | 39 | 20 | 32 | 50 | 56 | 73 | 93 |
| | | Alt | $1 - \beta_l = 0.977$ | 49 | 28 | 40 | 61 | 66 | 86 | 104 |
| 1/64 | 64 | Null | $\alpha_l = 0.012$ | 47 | 25 | 38 | 58 | 64 | 84 | 102 |
| | | Alt | $1 - \beta_l = 0.988$ | 50 | 28 | 40 | 62 | 68 | 86 | 104 |

**Table 5**

*Simulation results for likelihood study design in illustrative example with $H_1 : \psi = 0.415$*
*($\Delta = 0.44$). The data were sequentially examined upon observing each event after a total of*
*10 events were observed. The maximum number of subjects per group was assumed to be 50.*

| Evidence | | Oper. Char. | | Non-stop | Distribution of stopping times (events) | | | | | | | |
|---|---|---|---|---|---|---|---|---|---|---|---|---|
| $k_0$ | $k_1$ | | | | $E[D]$ | 25% | 50% | 75% | 80% | 90% | 95% | 100% |
| 1/8 | 8 | Null | $\alpha_l = 0.094$ | 0.004 | 25 | 14 | 20 | 30 | 35 | 46 | 58 | 80 |
| | | Alt | $1 - \beta_l = 0.892$ | 0.026 | 23 | 13 | 17 | 29 | 32 | 40 | 53 | 71 |
| 1/20 | 20 | Null | $\alpha_l = 0.032$ | 0.056 | 36 | 20 | 29 | 45 | 52 | 67 | 75 | 87 |
| | | Alt | $1 - \beta_l = 0.856$ | 0.112 | 33 | 20 | 29 | 42 | 47 | 58 | 62 | 71 |
| 1/32 | 32 | Null | $\alpha_l = 0.016$ | 0.096 | 40 | 23 | 34 | 53 | 60 | 72 | 77 | 87 |
| | | Alt | $1 - \beta_l = 0.824$ | 0.148 | 37 | 25 | 34 | 49 | 52 | 61 | 64 | 71 |
| 1/64 | 64 | Null | $\alpha_l = 0.004$ | 0.138 | 45 | 27 | 40 | 63 | 68 | 76 | 78 | 87 |
| | | Alt | $1 - \beta_l = 0.758$ | 0.224 | 43 | 31 | 41 | 55 | 58 | 62 | 65 | 71 |